\documentstyle[11pt,aaspp4]{article}
\def\gax{\mathrel{\raise.3ex\hbox{$>$}\mkern-14mu\lower0.6ex\hbox{$\sim$}}}
\def\lax{\mathrel{\raise.3ex\hbox{$<$}\mkern-14mu\lower0.6ex\hbox{$\sim$}}}
\def\gtorder{\mathrel{\raise.3ex\hbox{$>$}\mkern-14mu
             \lower0.6ex\hbox{$\sim$}}}
\def\ltorder{\mathrel{\raise.3ex\hbox{$<$}\mkern-14mu
             \lower0.6ex\hbox{$\sim$}}}

\begin{document}

\title{What Do Gravitational Lens Time Delays Measure?}

\author{C.S. Kochanek}
\affil{Harvard-Smithsonian Center for Astrophysics, 
       60 Garden Street, Cambridge, MA 02138}
\affil{email: ckochanek@cfa.harvard.edu}

\def\kbar{\langle \kappa \vphantom{R_1^2}\rangle}
\def\kbarp{\langle \kappa \vphantom{R_1^2}\rangle}
\def\ra{R_1}
\def\rb{R_2}
\def\rbar{\langle R \rangle}
\def\dr{\Delta R}

\begin{abstract}
Gravitational lens time delays depend on the Hubble constant, the observed
image positions, and the surface mass density of the lens in the annulus 
between the images.  Simple time delay lenses like PG1115+080, SBS1520+530, 
B1600+434, PKS1830--211 and HE2149--2745 have 
$H_0 = A(1-\kbar)+B\kbar(\eta-1)$ where the two coefficients
$A \simeq 90$~km/s~Mpc and $B \simeq 10$~km/s~Mpc depend on the
measured delays and the observed image positions, $\kbar$ is the
mean surface density in the annulus between the images, and 
there is a small correction from the logarithmic slope $\eta \simeq 2$
of the surface density profile, $\kappa \propto R^{1-\eta}$, in the 
annulus.  These 5 systems are very homogeneous, since for fixed 
$H_0=100h$~km/s~Mpc they must have the same surface density,
$\kbarp=1.11-1.22 h \pm 0.04$, with an upper bound of $\sigma_\kappa <0.07$
on any dispersion in $\kbar$ beyond those due to the measurement errors. 
If the lenses have their expected dark halos, $\kbar \simeq 0.5$ and
$H_0\simeq 51\pm 5$~km/s~Mpc, while if they have constant mass-to-light 
ratios, $\kbar \simeq 0.1$--$0.2$ and $H_0 \simeq 73\pm 8$~km/s~Mpc.
More complicated lenses with multiple components or strong perturbations
from nearby clusters, like RXJ0911+0551 and Q0957+561, are easily recognized 
because they have significantly different coefficients. 
\end{abstract}


\section{Introduction}

Gravitational lens time delay measurements can determine the Hubble constant
$H_0$ given a model for the gravitational potential of the lens galaxy
(Refsdal~\cite{Refsdal64}).  With 9 well-measured time delays (see Schechter~\cite{Schechter00}),
gravitational lenses have become a serious alternative to the local distance scale. 
Kochanek~(\cite{Kochanek02a}) modeled the 5 simple, well-observed 
time delay lenses to find that $H_0=48_{-4}^{+7}$~km/s~Mpc if the lenses have
isothermal mass distributions (flat rotation curves) and that $H_0=71\pm6$~km/s~Mpc
if they have constant mass-to-light ($M/L$) ratios.  Intermediate values
can be found by adjusting the dark matter halo between these two extreme
models, but the results agree with the local estimate of $H_0=72\pm8$~km/s~Mpc
by the HST Key Project (Freedman et al.~\cite{Freedman01}) only if the
lenses contain little or no dark matter. Models of
other lenses (e.g. Munoz, Kochanek \& Keeton~\cite{Munoz01}), local stellar 
dynamical measurements (e.g. Rix et al.~\cite{Rix97}, Romanowsky \& Kochanek~\cite{Romanowsky99},
Gerhard et al.~\cite{Gerhard01}, Treu \& Koopmans~\cite{Treu02}), weak lensing 
(e.g. Guzik \& Seljak~\cite{Guzik02}) and X-ray (e.g. Fabbiano~\cite{Fabbiano89},
Lowenstein \& White~\cite{Lowenstein99}) measurements all require mass distributions
close to the dark matter limit.

The dependence of the estimate of $H_0$ from a gravitational lens time delay 
on the mass distribution of the lens is well known, both from models of particular 
time delay lenses (e.g. Keeton \& Kochanek~\cite{Keeton97}, Impey et al.~\cite{Impey98},
Koopmans \& Fassnacht~\cite{Koopmans99}, Lehar et al.~\cite{Lehar00}, 
Keeton et al.~\cite{Keeton00}, Williams \& Saha~\cite{Williams00}, Winn et al.~\cite{Winn02}) 
and from general analytic principles (Falco et al.~\cite{Falco85}, 
Gorenstein et al.~\cite{Gorenstein88}, Refsdal \& Surdej~\cite{Refsdal94},
Witt, Mao \& Schechter~\cite{Witt95}, Witt, Mao \& Keeton~\cite{Witt00},
Schechter~\cite{Schechter00}, Rusin~\cite{Rusin00}, Saha~\cite{Saha00}, Zhao \& Pronk~\cite{Zhao01}, 
Wucknitz~\cite{Wucknitz02}, Oguri et al.~\cite{Oguri02}).  The most important trend 
is that the predicted time delays, or the inferred $H_0$, increase as the mass 
distribution becomes more centrally concentrated.  In particular, 
Falco et al.~(\cite{Falco85}) showed that adding a constant surface
density sheet to the mass distribution has no observable effect other than 
to rescale the time delay (the mass sheet degeneracy), and  
Witt et al.~(\cite{Witt00}) showed that the Hubble constants
estimated from lens potentials of the form $\phi \propto R^{3-\eta}$
($\rho \propto r^{-\eta}$) roughly scale as $H_0 \propto (\eta-1)/\Delta t$, almost
independent of the angular structure of the potential.  

Our objective in this paper is to clarify the physical properties of the lenses
which control the expected time delays after making full use of the astrometric 
constraints available for a typical lens.  We will show that time delays are
controlled by a local property of the lens, the average surface density in the 
annulus between the lensed images.  Reducing the model-dependence of the time
delay estimates to such a simple physical property of the lens leads to
simple, accurate scaling laws for $H_0$,  provides model-independent
tests for the homogeneity of the time delay lenses, demonstrates that standard
parametric models have the necessary degrees of freedom to study degeneracies
in estimates of $H_0$ from time delay measurements, and makes it easy to 
theoretically predict time delays for standard halo models.
In \S2 we develop analytic results
for lenses in arbitrary spherical density distributions.  We first consider
circular lens potentials (\S2.1) and then examine the effects of adding 
an external shear (\S2.2), the quadrupole of a singular isothermal 
ellipsoid (\S2.3) or a general quadrupole (\S2.4).  These analytic results
show that the time delays are largely determined by the mean surface
density $\kbar=\langle\Sigma\rangle/\Sigma_c$ in the annulus between
the images.  In \S3 we show that the $H_0$ estimates for most
lenses can be reduced to a linear function of $\kbar$, and that
these simple scaling solutions reproduce the full numerical models by 
Kochanek~(\cite{Kochanek02a}).  In \S4 we discuss the consequences of these
results for the homogeneity of the lens population.

\def\bfx{{\bf x}}
\def\bfu{{\bf u}}
\def\grad{{\bf \nabla}}
\def\ka{\kappa_1}
\def\kb{\kappa_2}

\section{Analytic Models}

We start from the dimensionless time delay (see Schneider,
Ehlers \& Falco~\cite{Schneider92}),
\begin{equation}
   \tau = {1 \over 2}\left( x -u \right)^2 + {1 \over 2}\left( y - v \right)^2 -
    \phi(x,y)
   \label{eqn:delay}
\end{equation}
where $x$ and $y$ are the angular coordinates on the lens plane, $u$ and
$v$ are angular coordinates on the source plane, and $\phi(x,y)$ is the
lens potential.  The center of the lens galaxy is used as the
coordinate origin.  The true delay is
\begin{equation}
   t = { 1+z_l \over H_0 } {D_{OL} D_{OS} \over r_H D_{LS}} \tau 
          \equiv \hat{D}_{eff} H_0^{-1} \tau  
    \label{eqn:physdelay}    
\end{equation}
where the $D_{ij}$ are angular diameter distances between the observer,
lens and source, $z_l$ is the lens redshift, and $H_0$ is the
Hubble constant.\footnote{We note in passing that the distance factor
scaling of the time delays is simpler if comoving distances are used
rather than angular diameter distances.  The extra factor of $1+z_l$
is absorbed into the distances. Comoving distances, not angular 
diameter distances, are the natural cosmological distances for 
gravitational lensing calculations as they almost always lead to
simpler analytic expressions.  }  
The factor containing the distances and the Hubble radius,
$r_H=c/H_0$, depends on the cosmological model, the source redshift
$z_s$ and the lens redshift $z_l$, but not the Hubble constant.
We assume a $\Omega_0=0.3$ flat universe, but the cosmological
dependence of the results is very weak.  The lens potential
satisfies the Poisson equation $\nabla^2 \phi = 2 \kappa$ where
$\kappa=\Sigma/\Sigma_c$ is the surface density of the lens in
units of the critical surface density $\Sigma_c=c^2 D_{OS}/4\pi G D_{OL}D_{LS}$.
The first two terms of the time delay (Eqn.~\ref{eqn:delay}) form
the geometric delay from the bending
of the rays, and the remaining term is the Shapiro delay from
passing through the gravitational potential of the lens.  We
observe images at the solutions of $\grad \tau=0$.  

We derive the time delay differences for a pair of images in a series of 
analytic models based on multipole expansions for the lens potential
(see Kochanek~\cite{Kochanek91} and Trotter, Winn \& Hewitt~\cite{Trotter00}
for other applications of multipole methods to gravitational lenses).
Our objective is to achieve expressions for the delays which are 
accurate to $\ltorder 5\%$  when estimating the Hubble constant.
Two images at radius $\ra$ and $\rb$ have an average radius of 
$\rbar=(\ra+\rb)/2$ and bracket an annulus of width $\dr=\ra-\rb$.
For the monopole of the lens, our accuracy goal will require expansions
of order $(\dr/\rbar)^3$.  The ellipticity of the lens potential
generally satisfies $\epsilon < \dr/\rbar$ for two-image lenses
and $\epsilon \sim \dr/\rbar$ for four-image lenses.  Assuming a
standard multipole sequence for a roughly ellipsoidal mass distribution,
the quadrupole ($\exp(2 i \theta)$) is of order $\epsilon$ and
can be treated to one lower order of expansion than the monopole,
and higher order poles ($\exp( 2 m i \theta)$) can be neglected since 
they are of order $\epsilon^m$.   
We start in \S2.1 by estimating the time delays for an
arbitrary spherical lens.  Next we add a quadrupole potential to
the lens:  in \S2.2 we consider an external (tidal) shear, in \S2.3
we consider the quadrupole produced by a singular isothermal ellipsoid,
and in \S2.4 we consider a quadrupole with an arbitrary ratio between
the internal and external contributions to the quadrupole near the images.

\subsection{Spherical Models}

We start with a lens having two images on opposite sides of a spherical
lens, $\phi(x,y)=\phi_0(R)$, defined by a monopole potential $\phi_0(R)$.
The outer image, at radius $\ra$, is at a minimum of the time delay 
surface and the inner image, at radius $\rb$, is at a saddle point,
and the two images and the lens lie on a line. Starting from Eqn.~(\ref{eqn:delay}),  
the geometric delay between the two images can be written 
\begin{equation}
    \Delta\tau_{12,geom} = 
     -{ 1 \over 2 } \left( \ra^2-\rb^2 \right) \left( 1-2\kbar \right)
\end{equation}
where 
\begin{equation}
     \kbar = { 2 \int_{\ra}^{\rb} \kappa(u) u du \over \rb^2-\ra^2 }
\end{equation}
is the average surface density in the annulus bounded by the images
in units of the critical surface density.  Only the derivative of the
potential is needed to determine the geometric delay, and the 
mean surface density enters because the radial derivatives $\phi_0'$
of the monopole at the positions of the two images are related by
\begin{equation}
      R_2 \phi_0'(R_2) = R_1 \phi_0'(R_1) + \kbar (R_2^2-R_1^2)
\end{equation}
independent of the density distribution in the annulus.  For an isothermal lens
$\kbar=1/2$ and the geometric contribution to the delay is identically
zero for all images.  The Shapiro delay difference is 
more complicated,
\begin{equation}
   \Delta\tau_{12,Shap} = { 1 \over 2 } \kbar \left( \rb^2-\ra^2 \right) +
     \ra \rb \left( 1-\kbar \right) \ln { \rb \over \ra} -
     2\int_{\ra}^{\rb} u du \left( \kappa(u)-\kbar \right) \ln {u\over \rb}, 
     \label{eqn:shapdelay}
\end{equation}
because the Shapiro delay depends on the surface density distribution in the
annulus as well as the mean surface density.   The first term, which depends
only on $\kbar$, is the most important, as the contribution from the integral
is smaller by $(\dr/\rbar)^2$.  Combining the terms,
the time delay between the images,
\begin{equation}
   \Delta\tau_{12}=\tau_1-\tau_2 =
     \left( 1-\kbar \right) 
  \left[ -{ 1 \over 2 }\left( \ra^2-\rb^2 \right)  + \ra \rb \ln { \rb \over \ra}\right]
      - 2\int_{\ra}^{\rb} u du \left( \kappa(u)-\kbar \right)  \ln { u \over \rb },
   \label{eqn:gensphere}
\end{equation}
depends only on the image positions and the surface density of the
lens in the annulus between them (Gorenstein et al.~\cite{Gorenstein88}).  
The first term is the same as the time
delay of a point mass lens combined with a uniform sheet of density $\kbar$.

We can capture the essential elements of Eqn.~(\ref{eqn:gensphere}) without the
complexity by assuming that the surface density in the annulus between the
images is a power law $\kappa=\ka(R/\ra)^{1-\eta}$ and then expanding the
time delay as a series in the ratio of the thickness of the annulus $\dr=\rb-\ra$
to its average radius $\rbar=(\ra+\rb)/2$ to find that
\begin{equation}
   \Delta\tau_{12} = - \left( \ra^2-\rb^2 \right) 
           \left[ \left( 1-\kbar \right) - 
       { 1 -\eta \kbar \over 12 } \left( { \dr \over \rbar } \right)^2
           + O\left( \left( { \dr \over \rbar } \right)^4\right) \right].
    \label{eqn:delaymod0}
\end{equation}
The first term is the exact delay for a singular isothermal sphere (SIS, $\eta=2$, 
$\kbar=1/2$), and the series converges rapidly even for large values of $\dr/\rbar$.
For example, the first term approximates the time delay of a point mass
lens with only a 9\% error even as $\dr/\rbar \rightarrow 1$.  The second order
correction adjusts for the small changes in the critical radius (inside which the
average surface density is unity) with the shape of the density profile.  It
comes only from the Shapiro delay (Eqn.~\ref{eqn:gensphere}) and can be 
decomposed ($1-\eta\kbar = (1-\kbar)+\kbar(\eta-1)$) into a contribution
depending only on the average surface density and a contribution which depends
on its distribution.  The dependence on the logarithmic slope $\eta$ arises 
only from the integral term of the Shapiro delay.  The scaling
observed by Witt et al.~(\cite{Witt00}) for potentials of the form $\phi \propto R^{3-\eta}$
arises because $1-\kbar=(\eta-1)/2$ near the critical line of the lens. 

\subsection{A Spherical Lens in an External Shear}

Real lenses are not spherical, so we next consider a range of models with a quadrupole
as well as a monopole.  The monopole depends only on the total mass inside the Einstein
ring and the surface density in the annulus.  The quadrupole depends on the total
internal quadrupole (due to the quadrupole moment of the mass inside the Einstein ring), 
the total external quadrupole (due to the quadrupole moment of the mass outside the ring) 
and the quadrupole component from the surface density inside the annulus.  Because the
ellipticity of the potential is small ($\ltorder \dr/\rbar$) we can treat the quadrupole 
to a lower order of expansion than the monopole and ignore the quadrupole component
of the surface density in the annulus.  For our analytic results, but not for the
numerical results in \S3, we assume that the internal and external quadrupoles are 
aligned.  Thus, we need three numbers, two amplitudes and an orientation, to define
the quadrupole.  Higher order angular structures (octopole etc.) will be smaller than
the quadrupole by one or more powers of the ellipticity of the potential, and can
be neglected.

We start with two familiar examples of quadrupoles.  We first consider an external
(tidal) shear, where the quadrupole is generated entirely by material outside the
region with the images.  In \S2.3 we discuss the quadrupole produced by a 
singular isothermal ellipsoid. Then in \S2.4 we discuss the general aligned
quadrupole.  Ideally, we would present simple results to the same order ($(\dr/\rbar)^3$)
as we did for the monopole.  However, our objective is to illustrate the effects of 
quadrupole structures on time delays, so we will not reproduce terms which are grotesquely 
complicated  simply to reach an arbitrary order of expansion.

\def\dthet{\Delta\theta_{12}}
\def\sthet{\theta_1+\theta_2}
\def\tg{\theta_{ext}}

The quadrupole potential associated with an external shear is 
$\phi_2=(\gamma_{ext}/2) R^2 \cos 2(\theta-\theta_{ext})$,
where $\gamma_{ext}$ is the amplitude of the external shear and $\theta_{ext}$
is its orientation.  As in \S2.1, the monopole is defined by the mass inside the
Einstein ring, the average surface density $\kbar$ in the annulus and
the logarithmic derivative of the surface density $\eta$.  We model
two images located at ($\ra\cos\theta_1$,$\ra\sin\theta_1$) and
($\rb\cos\theta_2$,$\rb\sin\theta_2$) from the lens center.  
Using the positions of two images we solve for the mass inside
the Einstein ring and the amplitude of the external shear.
This results in expressions which depend on the surface density in
the annulus ($\kbar$, $\eta$) and the orientation of the shear ($\theta_{ext}$).  
The shear axis seems the best choice for an independent variable in our analytic 
models because its value is little affected by
changes in the monopole or quadrupole structure of the lens 
(see Kochanek~\cite{Kochanek91}). 
If we again expand the delay as a series in $\dr/\rbar$,
\begin{equation}  
      \Delta\tau_{12} = - \left( \ra^2-\rb^2 \right) \left[ T_0 + {\dr\over\rbar} T_1 +
              \left( {\dr\over\rbar} \right)^2 T_2 \cdots \right]   
       \label{eqn:expansion}
\end{equation}
then the leading term,
\begin{equation} 
    T_0 = \left(1-\kbar\right) \sin^2\left({ \dthet \over 2 } \right) 
      \longrightarrow \left(1-\kbar\right) \quad\hbox{as}\quad \dthet\rightarrow\pi,
     \label{eqn:a0term}
\end{equation}
changes from the result for a spherical system only in adding a dependence on the 
angle between the images $\dthet=\theta_1-\theta_2$.  
The limit as
$\dthet\rightarrow\pi$ shows that we recover the result for a circular lens and
images collinear with the lens (Eqn.~\ref{eqn:delaymod0}).  The first order correction,
\begin{equation}
     T_1 =   - { 1 \over 2 } \left(1-\kbar\right) \sin\left(\dthet\right)\cot(\sthet-2\tg)
        \longrightarrow 0 \quad\hbox{as}\quad \dthet\rightarrow\pi,
\end{equation}
depends on the shear axis $\tg$, but the second order correction,
\begin{equation}
    T_2 =  { 1 \over 12 }\left[ -\left(1-\kbar\right)\left( 4+3\cos\dthet\right) 
                + \kbar \left( \eta-1 \right) \right]
       \longrightarrow - { 1 - \eta \kbar \over 12 } \quad\hbox{as}\quad \dthet\rightarrow\pi,
\end{equation}
depends only on the surface density in the annulus.  Note that the higher
order terms return to the spherical case for images collinear with the lens.
The amplitude of the shear is
\begin{equation}
    \gamma_{ext} = -{ \left(1-\kbar\right)\left( \ra^2-\rb^2 \right) \over 
        \left(\ra^2 +\rb^2 -2\ra\rb\cos\dthet \right)}
                  { \sin \dthet \over \sin \left(\sthet-2\tg \right) }
        \longrightarrow 0 \quad\hbox{as}\quad \dthet\rightarrow\pi,
      \label{eqn:shear}
\end{equation}
so the singularity in $T_1$ when $\sthet-2\tg\rightarrow 0$ is due to a singularity in
the shear that will not arise in practice since $\tg$ must be set using the parameters
of a reasonable macro model. Also note that the amplitude of the shear scales
as $1-\kbar$, so that lenses with lower surface densities require larger shears.  

The dependence of the delay on the structure of the surface density in the annulus
is identical to that for the spherical lens, with the integral in the Shapiro 
delay (Eqn.~\ref{eqn:shapdelay}) adding a term $(\eta-1)\kbar/12$ to the
$T_2$ coefficient. All other terms in the time delay depend only on the 
derivatives of the monopole and are functions only of $\kbar$ even for an arbitrary 
monopole surface density.  The term vanishes for $\eta=1$ when 
$\kappa =\kappa_1(R/R_1)^{1-\eta}$ becomes a constant surface density.

\subsection{A Spherical Lens in an SIE Quadrupole}

Unlike our result for an external shear, where the time delay depends on the
angle between the images even at the lowest order of the expansion, 
Witt et al.~(\cite{Witt00}) found that the time delays for potentials of the 
form $\phi \propto R {\cal F}(\theta)$ (which includes the standard singular
isotherma ellipsoid or SIE potential) depend only on the image radii.
These models have a quadrupole potential of the form
$\phi_2 = \epsilon R \rbar \cos 2(\theta-\theta_\epsilon)$.
The leading term in the time delay,
\begin{equation}
     T_0 = 1-\kbar
\end{equation}
is independent of the angle, just as for the Witt et al.~(\cite{Witt00}) potential.
The second term is a correction for the effects of the ellipticity,
\begin{equation}
     T_1 = -{1\over 8}\left(1-\kbar\right) 
              \left[ 5 \cos\left( \dthet/2\right) + \cos \left(3\dthet/2\right) \right] 
              \cot\left(\sthet-2\tg\right) \csc^3\left(\dthet/2\right)
\end{equation}
The next term is too complicated to present in full, but it has the structure
\begin{equation}
    T_2 = \left(1-\kbar\right)g(\theta_1,\theta_2,\theta_\epsilon)+ {1\over 12}(\eta-1)\kbar.
\end{equation}
The first factor depends on a very complicated angular function, and the second factor
is the correction to the Shapiro delay from the logarithmic slope $\eta$ of the monopole 
surface density.  To lowest order, the ellipticity of the potential is
\begin{equation}
   \epsilon \simeq - { 1-\kbar \over 4 } {\dr \over \rbar} \sin \dthet \csc^4 (\dthet/2)
         \csc(\sthet-2\theta_\epsilon).
      \label{eqn:ellips}
\end{equation}
We again find that the ellipticity scales as $1-\kbar$ and that the singularities in
$T_1$ correspond to singularities in $\epsilon$.

\subsection{A Spherical Lens in a General Quadrupole}

The quadrupoles of an external shear and an SIE differ in the balance between 
internal and external quadrupoles.  Since they also produce different angular
dependences for the leading term in the time delay, it suggests that the 
angular dependence of the time delay is determined by the relative strength
of the two shear components.  We can test this by determining the leading
term in the time delay for a quadrupole which is an arbitrary sum of an
internal quadrupole of amplitude $\gamma_{int}$ and external quadrupole of
amplitude $\gamma_{ext}$, 
\begin{equation}
  \phi_2={ 1\over 2}\left(\gamma_{ext} R^2 + \gamma_{int} { \rbar^4 \over R^2} \right)
       \cos 2(\theta-\theta_\gamma).
\end{equation}
If we use the image positions to determine the total shear,
$\Gamma=\gamma_{int}+\gamma_{ext}$, as a function of the shear angle $\theta_\gamma$
and the ratio $f_{int}=\gamma_{int}/\Gamma$, we find that the leading term,
\begin{equation}
   T_0 = - \left(1-\kbar\right) { \sin^2\left(\dthet/2\right) \over 
           1 - 4 f_{int} \cos^2\left(\dthet/2\right) },
   \label{eqn:quaddep}
\end{equation}
has the standard scaling with $1-\kbar$ and an angular dependence determined
by the balance between the internal and external quadrupoles. For
an external shear ($f_{int}=0$) the denominator is unity, and for the
quadupole of an SIE ($f_{int}=1/4$ for $\kappa_2 \propto R^{-1}$) the  
numerator and denominator cancel to make $T_0$ independent of 
$\dthet$ just as in the Witt et al.~(\cite{Witt00}) results.  For
$f_{int} > 1/4$, where the internal shear begins to dominate, there are
angles where the expansion fails.  The higher order terms are 
too complex to be worth presenting for general configurations.  However, for
two-image lenses where the images lie on opposite sides of the 
lens, $\theta_2=\theta_1+\pi +\delta\theta$, with a small 
angular offset defined by $\xi=\delta\theta \rbar/\dr \ltorder 1$,
the delay depends little on the
structure of the quadrupole.  The expansion terms,
\begin{equation}
     T_0 = 1-\kbar, \qquad
     T_1 = 0 \quad\hbox{and}\quad
     T_2 = - { 1\over 12}(1-\eta)\kbar + {1\over 4}\xi \left(1-\kbar \right)
            \left[ 2\cos 2\theta_\gamma + \xi ( 4 f_{int}-1 ) \right]
\end{equation}
are nearly identical to those for a spherical lens (Eqn.~\ref{eqn:delaymod0})
because the quadrupole structure enters only at second order.

\section{A Simple Semi-Analytic Model}

\def\rbar{\langle R\rangle}

These analytic results suggest a simple semi-analytic model for understanding the 
time delays observed in gravitational lenses.  It cannot be a completely analytic
model because we require a numerical solution using all the lensed images to 
determine the quadrupole structure of the four-image lenses.  We continue to  
model the monopole by the mass inside
the average image radius and a power law, $\kappa(R)\propto R^{1-\eta}$,
for the surface density between the inner and outer images parameterized by
the mean surface density $\kbar$ and the logarithmic slope $\eta$.  The 
resulting expression for the monopole deflection,
\begin{equation}
     \phi_0'(R) =  b_0 { \rbar \over R } +  { 2 \over R } \int_{\rbar}^R \kappa(R)RdR,
\end{equation}
where $b_0 \sim \rbar$ determines the mass inside $\rbar$ is very similar to an early 
model of Q0957+561 by Borgeest \& Refsdal~(\cite{Borgeest84}) which combined a point mass 
and an isothermal sphere ($\eta=2$). Since many of the
lenses have $\dr/\rbar \sim 0.5$ we expanded the monopole to first order in the 
deviation of the density exponent from isothermal, $\delta\eta=\eta-2$, rather than
$\dr/\rbar$.  This leads to results that are exact for $\eta=2$ independent of the
width of the annulus, and are reasonably accurate for $1 \ltorder \eta \ltorder 3$. 
We combined the monopole with a general quadrupole
\begin{equation}
   \phi_2(R,\theta) = { 1 \over 2 } \gamma_{ext} R^2 \cos 2 (\theta-\theta_{ext}) +
            { 1 \over 2 } \gamma_{int} { \rbar^4 \over R^2 } \cos 2(\theta-\theta_{int})
\end{equation}
where the internal and external components have amplitudes and orientations of
($\gamma_{int}$,$\theta_{int}$) and ($\gamma_{ext}$,$\theta_{ext}$) respectively.  For 
each lens we determined the values of $b_0$ and the shear components which minimized 
the differences in the projected source positions, leaving expressions depending
only on the monopole structure in the annulus ($\kbar$ and $\eta$).  For the 
current time delay lenses, we cannot determine $\kbar$ or $\eta$ from the 
available constraints.  By doing the fits on the source plane, the solutions can be 
found rapidly using {\it mathematica} programs.  We did not fit the image 
flux ratios.  For the two-image lenses, where there are too few constraints to
determine all 5 variables, we stabilized the solutions by finding the minimum
shear solutions.\footnote{For two-image lenses we added on weak prior on the
shear amplitudes to the $\chi^2$ for the mapping of the images back to
a common source. The prior was defined by ${\bf \Gamma}\cdot {\bf \Gamma} \gamma_0^{-2}$ where
${\bf \Gamma}=(\gamma_{int}\cos 2\theta_{int},\gamma_{int}\sin 2\theta_{int},\gamma_{ext}\cos 2\theta_{ext},\gamma_{ext}\sin 2\theta_{ext})$ is the vector of Cartesian shear components
and we set $\gamma_0=0.05$. }
We experimented with the structure of the shear terms (and the shear prior for
the two-image lenses)
to ensure that the results
reported for $H_0$ are not affected by the assumed quadrupole structure.  For the
two-image lenses the results had no significant dependence on the quadrupole
structure (as expected from \S2.4), and for the four-image lenses the quadrupole
structure was well-determined by the image positions.  Errors
were estimated by Monte Carlo simulations and for $H_0$ they are dominated by the
errors in the time delay measurements.  

Once we have solved for the model parameters required to match the image positions,
we derive the time delays between images as a function of the remaining parameters
$\kbar$ and $\eta$.  Based on the analytic and numerical results, we can
express the results as simple scaling relations for the Hubble constant of the form 
\begin{equation}
    H_0 \simeq  A(1-\kbarp) + B \kbarp (\eta-1) + C
    \label{eqn:dimendelay}
\end{equation}
where the three coefficients ($A$, $B$ and $C$) have the units of km/s~Mpc and
scale inversely with the measured time delays.  We have isolated the terms more
physically here than in the expansions of \S2, isolating the dependence of the
logarithmic slope $\eta$ in the $B$ coefficient, but the relationship between the
coefficients and the expansion can be worked out for each model.  For example,
the $T_0$ term defined in Eqn.~(\ref{eqn:expansion}) contributes 
$ -(\ra^2-\rb^2) T_0 \hat{D}_{eff}/\Delta t_{12}$  to the $A$ coefficient,
where $\Delta t_{12}$ is the measured delay and $\hat{D}_{eff}$ is the
effective distance defined in Eqn.~(\ref{eqn:physdelay}).  In the analytic
models $C\equiv 0$. For simple lenses the
Hubble constant $H_0 \simeq A$ for very centrally concentrated lenses
($\kbarp\rightarrow 0$), and $H_0 \simeq (A+B)/2$ for an isothermal
model ($\kbarp=1/2$, $\eta=2$).  We order the images so that the time
delay is positive and $A>0$.  The effects of an additional mass sheet of density
$\kappa_{ext}$ can be easily added to the models.  The adjustments depend
on whether we regard $\kbarp$ as a fixed number or a fit parameter
whose value is also changed by the addition of the extra convergence.
If $\kbarp$ is a fixed number, then the estimate of the Hubble 
constant changes to
\begin{equation}
    H_0 \simeq  \left(1-\kappa_{ext}\right)\left[ A\left(1-\kbarp\right)+C \right] + B \kbarp (\eta-1).
\end{equation}
In most cases, however, we start with a lens whose mass and surface density
are scaled to fully explain the image splitting.  With the addition of an
external convergence, the annular surface density $\kbarp_0$ of this 
reference model is reduced to $\kbarp=(1-\kappa_{ext})\kbarp_0$ because
the extra convergence reduces the mass of the primary lens.  Thus,
if we specify $\kbarp_0$, the estimate of the Hubble constant with
the addition of an external convergence is 
\begin{equation}
    H_0 \simeq  (1-\kappa_{ext})\left[ A\left(1-\kbarp_0(1-\kappa_{ext})\right) + B \kbarp_0 (\eta-1)+C\right]
\end{equation}
and it has the familiar $H_0 \propto (1-\kappa_{ext})$ scaling of the mass-sheet 
degeneracy (Falco et al.~\cite{Falco85}).

\def\hd{\hphantom{1}}
\def\hm{\hphantom{-}}
\def\m1#1{\multicolumn{1}{c}{#1}}

\begin{deluxetable}{ccccccccr}
\tablecaption{Hubble Constant Models}
\tablewidth{0pt}
\tablehead{
  Lens        &Images &Delay  &$R_e/\rbar$ &$\dr\over2\rbar$ &A$_0$     &A         &B       &\m1{C} \nl
              &       &(days) &            &                 &$[H_0]$   &$[H_0]$   &$[H_0]$ &\m1{$[H_0]$ }
  }
\startdata
RXJ0911+0551  &AD     &$   146\pm\hd 4$  &$0.55$ &$0.58$ &$   272\pm   14$  &$\hd 82\pm\hd 5$  &$   69\pm 4$ &$    -9\pm 1$\nl
              &BD     &$              $  &       &$0.54$ &$   280\pm   14$  &$\hd 81\pm\hd 5$  &$   69\pm 5$ &$    -9\pm 1$\nl
              &CD     &$              $  &       &$0.60$ &$   262\pm   14$  &$\hd 82\pm\hd 6$  &$   68\pm 5$ &$    -9\pm 1$\nl
Q0957+561     &       &$   417\pm\hd 3$  &$0.63$ &$0.67$ &$   249\pm\hd 2$  &$   221\pm\hd 2$  &$   51\pm 1$ &$    -7\pm 0$\nl
PG1115+080    &A$_1$B &$\hd 12\pm\hd 2$  &$0.41$ &$0.10$ &$\hd 81\pm   17$  &$\hd 70\pm   14$  &$\hd 0\pm 0$ &$\hm  1\pm 0$\nl
              &A$_1$C &$\hd 13\pm\hd 2$  &       &$0.10$ &$   126\pm   22$  &$   112\pm   20$  &$\hd 9\pm 2$ &$\hm  0\pm 0$\nl
              &BC     &$\hd 25\pm\hd 2$  &       &$0.19$ &$\hd 97\pm\hd 8$  &$\hd 93\pm\hd 8$  &$\hd 5\pm 0$ &$\hm  0\pm 0$\nl
SBS1520+530   &       &$   129\pm\hd 3$  &$0.51$ &$0.52$ &$\hd 98\pm\hd 5$  &$\hd 92\pm\hd 5$  &$   10\pm 1$ &$    -1\pm 0$\nl
B1600+434     &       &$\hd 51\pm\hd 2$  &$0.94$ &$0.64$ &$   119\pm   14$  &$   104\pm   10$  &$   21\pm 9$ &$    -3\pm 2$\nl
PKS1830--211  &       &$\hd 26\pm\hd 4$  &       &$0.19$ &$\hd 85\pm   15$  &$\hd 88\pm   16$  &$\hd 1\pm 0$ &$\hm  0\pm 0$\nl
HE2149--2745  &       &$   103\pm   12$  &$0.55$ &$0.59$ &$\hd 95\pm   12$  &$\hd 84\pm   11$  &$   14\pm 2$ &$    -1\pm 0$\nl
\enddata
\tablecomments{
  For each lens and image pair we give the measured time delay in days, followed
  by the coefficients which determine the Hubble constant given the surface 
  density model.  We first give the single coefficient $A_0$ needed for the
  lowest order expansion $H_0 =A_0(1-\kbarp)$ for an arbitrary monopole in
  an external shear, and then the three coefficients $A$, $B$ and $C$ used in our 
  standard expansion (Eqn.~{\protect\ref{eqn:dimendelay}}) for an arbitrary
  monopole and quadropole.  The formal errors include the uncertainties in
  both the astrometry and time delays but are dominated by the time delays.
  This means that the uncertainties are almost perfectly correlated!
  We broaden the time delay errors to a minimum error of 5\% to account for
  systematic uncertainties such as convergence fluctuations from large scale
  structure.  The ratio $R_e/\rbar$ is the ratio of the (monopole) half-light 
  radius to the average image radius, and the ratio $\dr/2\rbar$ is an estimate
  of the thickness of the annulus. 
  }
\end{deluxetable}

\def\hd{\hphantom{1}}
\begin{deluxetable}{cccccccccc}
\tablecaption{Comparisons to Full Numerical Models}
\tablewidth{0pt}
\tablehead{
  Lens &Image &\multicolumn{2}{c}{Full Model $H_0$}  
              &\multicolumn{2}{c}{$\kbar_0/\kappa_{ext}$}
              &\multicolumn{2}{c}{Lowest Order $H_0$} 
              &\multicolumn{2}{c}{Standard $H_0$} \nl
       & &$DM$ &$M/L$ &$DM$ &$M/L$ &$DM$ &$M/L$ &$DM$ &$M/L$
  }
\startdata
RXJ0911+0551  &AD     &$49\pm4$   &$67\pm4$  &$0.50/0.24$ &$0.11/0.31$  &$   129\pm\hd 7$  &$   174\pm\hd 9$ &$58\pm\hd 3$  &$\cdots $\nl  
              &BD     &           &          &$0.50/0.24$ &$0.10/0.31$  &$   132\pm\hd 7$  &$   180\pm\hd 9$ &$58\pm\hd 3$  &$\cdots $\nl  
              &CD     &           &          &$0.50/0.24$ &$0.12/0.31$  &$   123\pm\hd 7$  &$   166\pm\hd 9$ &$58\pm\hd 3$  &$\cdots $\nl  
PG1115+080    &A$_1$B &$46\pm4$   &$69\pm5 $ &$0.50/0.09$ &$0.18/0.16$  &$\hd 40\pm\hd 8$  &$\hd 58\pm   12$ &$35\pm\hd 7$  &$\hd 51\pm   10$\nl
              &A$_1$C &           &          &$0.50/0.09$ &$0.11/0.16$  &$\hd 62\pm   11$  &$\hd 96\pm   17$ &$60\pm   10$  &$\hd 87\pm   15$\nl
              &BC     &           &          &$0.50/0.09$ &$0.15/0.16$  &$\hd 48\pm\hd 4$  &$\hd 71\pm\hd 6$ &$48\pm\hd 4$  &$\hd 69\pm\hd 6$\nl
SBS1520+530   &       &$53\pm6$   &$73\pm5$  &$0.50/0.00$ &$0.20/0.00$  &$\hd 49\pm\hd 3$  &$\hd 79\pm\hd 4$ &$51\pm\hd 3$  &$\hd 77\pm\hd 4$\nl
B1600+434     &       &$59\pm6$   &$71\pm9$  &$0.50/0.05$ &$0.43/0.18$  &$\hd 60\pm\hd 7$  &$\hd 63\pm\hd 7$ &$59\pm\hd 7$  &$\hd 68\pm\hd 9$\nl
HE2149--2745  &       &$47\pm8$   &$66\pm8$  &$0.50/0.00$ &$0.22/0.00$  &$\hd 48\pm\hd 6$  &$\hd 74\pm\hd 9$ &$48\pm\hd 6$  &$\hd 70\pm\hd 9$\nl
\enddata
\tablecomments{
  Hubble constant estimates for dark matter ($DM$) and constant mass-to-light
  ratio ($M/L$) models of the density distribution.
  All estimates of the Hubble constant are in units of km/s~Mpc.  The estimates from
  full numerical models are the dark matter and constant $M/L$ models from
  Kochanek~(\protect\cite{Kochanek02a}).  The lowest order $H_0$ estimates use
  the lowest order expansion $H_0=A_0(1-\kbar)$, while the standard $H_0$ 
  estimates use Eqn.~({\protect\ref{eqn:dimendelay}}).  The $\kbar_0/\kappa_{ext}$
  columns give the mean surface density $\kbar_0$ of an unperturbed spherical 
  model and the surface density due to external perturbers $\kappa_{ext}$ in the
  full numerical models.  For the isothermal models we used $\eta=2$ while
  for the constant $M/L$ models we used $\eta=3$. 
  }
\end{deluxetable}

We computed these simple models for 7 of the 9 time delay lenses, excluding only
B0218+357 (where the lens position is poorly known, Lehar et al.~\cite{Lehar00}) 
and B1608+656 (where the lens consists of two interacting galaxies, 
Koopmans \& Fassnacht~\cite{Koopmans99}).  We consider RXJ0911+0551
(Hjorth et al.~\cite{Hjorth02}), Q0957+561 (Schild \& Thomson~\cite{Schild97},
Kundic et al.~\cite{Kundic97}, Bernstein et al.~\cite{Bernstein97},
Keeton et al.~\cite{Keeton00}), PG1115+080 
(Schechter et al.~\cite{Schechter97},
Barkana~\cite{Barkana97}, Impey et al.~\cite{Impey98}), SBS1520+530 (Burud~\cite{Burud02b}, 
Faure et al.~\cite{Faure02}), B1600+434 (Burud et al.~\cite{Burud00}, 
Koopmans et al.~\cite{Koopmans00}), PKS1830--211 (Lovell et al.~\cite{Lovell98},
Winn et al.~\cite{Winn02}, Courbin et al.~(\cite{Courbin02})) 
and HE2149--2745 (Burud et al.~\cite{Burud02}).  The system parameters
are generally derived from the CfA/Arizona Space Telescope Lens Survey
(CASTLES, Falco et al.~\cite{Falco01}) images of the systems.

Of these 7 systems, we expect our simple models to have difficulty with 
Q0957+561 and RXJ0911+0551. Q0957+561 is near the center of a 
cluster (Keeton et al.~\cite{Keeton00}, Chartas et al.~\cite{Chartas02}) 
where an external shear will poorly describe the effect of the
cluster on the image positions.  RXJ0911+0551 is on the outskirts of a 
cluster (see Morgan et al.~\cite{Morgan01}) and the primary lens 
also has a small satellite.   The external shear approximation is a
poor one for the perturbations from the RXJ0911+0551 cluster.  It is
at best a marginal approximation for $\kbar \sim 1/2$ and it fails
completely for $\kbar \sim 0$ (recall that $\gamma \propto (1-\kbarp)$, 
see Eqns.~\ref{eqn:shear} and \ref{eqn:ellips}).  
We included an SIS satellite in our model of RXJ0911+0551,
so it will match the Kochanek~(\cite{Kochanek02a}) dark matter models, but
it is difficult to treat the two potentials self-consistently within our
approximation.  We also note that our estimates for 
PKS1830--211 are invalid if we adopt the Courbin et al.~(\cite{Courbin02}) 
multiple-lens  interpretation of the system rather than the 
Winn et al.~(\cite{Winn02}) single-lens interpretation. 

We now check our approximations against full numerical models for 
RXJ0911+0551, PG1115+080, SBS1520+530, B1600+434 and HE2149--2745
by Kochanek~(\cite{Kochanek02a}).
We first checked the dependence of the results on the assumed quadrupole
structure by fitting models with either a pure external shear, an SIE 
quadrupole or a pure internal shear.  For the two-image lenses, the $H_0$
estimates are essentially independent of the quadrupole structure, as
expected from the analysis in \S2.4.  The largest fractional changes
in $H_0$ are approximately 3\%.  This is fortunate because the observational
constraints on the two-image lenses cannot distinguish between the quadrupole
structures. When we allow a general quadrupole, which matches realistic models 
where there are contributions to the angular structure from both tidal 
shears and the ellipticity of the lens galaxy (Keeton, Kochanek \& 
Seljak~\cite{Keeton97b}), all the doubles can be fit perfectly and
have $H_0$ estimates consistent with the three constrained shear
models.

For the two four-image lenses, RXJ0911+0551 and PG1115+080, the quadrupole
structure is enormously important because the functional dependence of
the delay on the angle between the images is so strong once the images
do not lie on opposite sides of the lens.  For example, for an SIS
monopole ($\kbar=1/2$, $\eta=2$), the $H_0$ estimates from the BC 
image pair in PG1115+080 are $47$ and $67$~km/s~Mpc for an
external shear and an SIE quadrupole respectively. The ratio
of the values matches that expected from Eqn.~(\ref{eqn:quaddep}).
Because they are four-image lenses, however, they also strongly
distinguish between quadrupole structures, with both lenses 
being far more consistent with an external shear than either a
SIE quadrupole or an internal shear. Time delay ratios are
also controlled by the quadrupole structure, and the measured
delay ratios in PG1115+080 are also more consistent with an
external shear than either an SIE quadrupole or an internal
shear.  The delay ratio seemed to be model independent in our studies
of PG1115+080 (Keeton \& Kochanek~\cite{Keeton97}, Impey et al.~\cite{Impey98})
because the models focused on changes in the monopole while always
optimizing the quadrupole structure.  Since the best fit quadrupole
was close to an external shear no matter what we used for the 
monopole structure, the time delay ratios showed almost no 
model dependence.  For models with a general
quadrupole, we find that $f_{int} \simeq 0.02$ and $0.01$ for
RXJ0911+0551 and PG1115+080 respectively. While very close to
a pure external shear ($f_{int}=0$), the general quadrupole
significantly improves the fit to the image positions.  

We adopt the general quadrupole models as our standard.  For the two-image
lenses the choice has negligible effect on the estimates of $H_0$, and for
the four-image systems it is required to provide an acceptable fit to the image
positions.  Table~1 presents the coefficients $A$, $B$, and $C$ (Eqn.~\ref{eqn:dimendelay})
as well as the coefficient $A_0$, with $H_0=A_0(1-\kbar)$, we would obtain 
using the lowest order term for an external shear model 
($T_0$ from Eqn.~\ref{eqn:a0term}).  The uncertainties in the coefficients
are almost entirely due to the time delay estimates, although they do
include the astrometric uncertainties in the image and lens positions.
These become important only for B1600+434 where the lens position is
relatively uncertain due to the dust lane in the galaxy (Koopmans, 
de Bruyn \& Jackson~\cite{Koopmans98}, Maller et al.~\cite{Maller00}). The
dominant role of the time delay uncertainties means that the 
uncertainties in the coefficients are almost perfectly correlated. 
With the exception of the two lenses in clusters, RXJ0911+0551 and
Q0957+561, the lowest order coefficient $A_0$ matches the higher
order estimate $A$ to approximately 10\%, the $B$ coefficient is
needed to achieve 10\% accuracy, and the $C$ coefficient is 
negligible compared to other sources of uncertainty.  The amplitude
of $B$ is largest for the lenses with thicker annuli
(larger $\dr/\rbar$).  In fact,
the coefficients clearly divide the sample into a set of five
very similar, relatively isolated lenses (PG1115+080, SBS1520+530,
B1600+434, PKS1830--211, and HE2149--2745) and the two lenses
in clusters (RXJ0911+0551 and Q0957+561).  

The real question, however, is whether these simple semi-analytic 
representations can reproduce the results of full numerical models.
Table~2 compares the semi-analytic estimates to the full numerical
lens models from Kochanek~(\cite{Kochanek02a}) for either a dark
matter dominated, SIE lens model ($\kbar_0=1/2$, $\eta=2$) or a
constant $M/L$ model where $\kbar_0\sim0.1$--$0.2$ for most lenses.
We defined $\kbar_0$ for the constant $M/L$ models by the average density 
in the annulus for a spherical model using the intermediate axis effective 
radius (the geometric mean of the major and minor axes) for the scale 
length in a de Vaucouleurs profile. For a lens with surface density
profile $\Sigma(R)$, the critical surface density profile is simply
\begin{equation}
   \kappa(R) =  { R_c^2 \Sigma(R) \over 2 \int_0^{R_c} R\Sigma(R) dR }
\end{equation}
where $R_c$ is the average critical radius of the lens.  It depends
only on the shape of the profile because for a spherical lens the
mean density inside the critical radius is equal to the critical
density so the image geometry determines the mass scale.  We also
include the effects of the external convergences $\kappa_{ext}$ from
the numerical models.  We do not include the constant $M/L$ 
semi-analytic estimates for RXJ0911+0551 where the treatment
of the nearby cluster as an external shear becomes untenable.
  
First, note that the estimates for each
image pair in the four-image lenses RXJ0911+0551 and PG1115+080 are 
mutually consistent.  This holds for the A$_2$B and A$_2$C pairs of
PG1115+080 as well. 
Second, for the simple lenses, the lowest order
and standard approximations are nearly identical for the dark matter
(high $\kbarp$) models, but the lowest order approximation tends to
give higher estimates for $H_0$ for the constant $M/L$ (low $\kbarp$)
models.  For the more complicated lens RXJ0911+0551, the differences
are enormous. Finally, the approximate solutions are an 
excellent match to the full numerical models using ellipsoidal lenses
in external shear fields.  Using the AD image pair from RXJ0911+0551 and
the BC image pair from PG1115+080, the average difference between
the approximate and numerical dark matter models is 
$\Delta H_0 = 2 \pm 4$~km/s~Mpc with almost all the differences due
to the most complex system, RXJ0911+0551.  The approximate model for
RXJ0911+0551 completely fails in the constant $M/L$ limit (see above), 
but if we compare the remaining four lenses the difference is 
$\Delta H_0= 2\pm 1$~km/s~Mpc.  For comparison, the average
differences for the lowest order approximation are
$\Delta H_0 = 0\pm 2$~km/s~Mpc and $\Delta H_0 = 2\pm 6$~km/s~Mpc
respectively, but we have to drop RXJ0911+0551 for the dark matter
models as well as the constant $M/L$ models.    Some of these residuals 
may be due to the ambiguities in defining $\kbar$ for ellipsoidal lens
galaxies.  We conclude that our approximate solutions generally provide
estimates for $H_0$ to better than 10\% individually and to
better than 5\% as an ensemble. 

\section{Discussion}

Gravitational lens time delays are determined by the Hubble constant, the
positions of the lensed images, and the surface density in the annulus
bounded by the images.  The average surface density $\kbar$ in the annulus
is more important than its distribution.  The relationship between time
delays and the local surface density is exact for circular lenses 
(Gorenstein et al.~\cite{Gorenstein88}).  While it is not exactly true
for non-circular lenses, it is true in practice.  In two-image lenses,
where the images lie on opposite sides of the lens galaxy, the delays
are insensitive to the angular structure of the lens.  In four-image
lenses the delays are very sensitive to the quadrupole structure of the
potential, but the image positions tightly constrain the quadrupole and
leave the surface density as the only important variable.  Our local
interpretation of time delays agrees with earlier understandings
based on global models and degeneracies, but better isolates the
relevant physical properties of lens models.  It also allows us to 
find simple scaling solutions for the Hubble constant which accurately
reproduce the results of full numerical models.  These scaling solutions
can be used to classify time delay lenses, since their structure differs
for simple lenses and lenses strongly perturbed by clusters, and to
quickly estimate the Hubble constant predicted for any assumed radial
mass distribution.

We can also use the scaling solutions to study the homogeneity of the lens
population.
Because $H_0$ must be a universal constant, we can use the requirement
that the lens all produce the same Hubble constant to estimate the surface
density differences between the lenses.  If we assume a universal density
slope $\eta$, then the five simple lenses (PG1115+080, SBS1520+530,
B1600+434, PKS1830--211 and HE2149--2745) are consistent with a common value
for the average surface density of 
\begin{equation}
   \kbarp \simeq 1 - 1.07 h + 0.14 (\eta-1)(1-h) \pm 0.04,
\end{equation}
where $H=100h$~km/s~Mpc.  Formally, only limited ranges for the Hubble 
constant are consistent with a universal surface density, with $h <0.7$, 
$h=0.5\pm0.6$ and $h=0.9\pm0.3$ for $\eta=1$ (constant surface density),
$2$ (isothermal) and $3$ respectively.  An alternative way to characterize
the similarities is use one lens as a reference and then estimate 
the surface density differences between the reference lens and the other systems.  
For example, if we assume $\eta=2$ and use PG1115+080 as the reference system, 
we find that $\kbarp=1.05-1.13h\pm0.04$ for PG1115+080.  The surface density 
differences are $\Delta\kappa=0.07-0.09h\pm0.06$, $0.19-0.07h\pm0.11$, $0.00-0.00h\pm0.19$
and $0.14-0.28h\pm0.15$ for SBS1520+530, B1600+434, PKS1830--211 and
HE2149--2745 respectively.  The images in PG1115+080, SBS1520+530 and HE2149--2745
all lie at approximately the same radius relative to the lens galaxy 
($R_e/\rbar \simeq 0.5$) while the images in B1600+434 are close to the
half-light radius ($R_e/\rbar \simeq 1$).
Reasonable slope differences between the lenses can change
the surface density differences by $0.00$--$0.27(1-h)\Delta\eta$ depending
on the lens.  At least for constant $\eta$, the upper bound on the existence
of any scatter in the surface density beyond that implied by the measurement
errors is $\sigma_\kappa \ltorder 0.07$, so these five simple lenses have
very homogeneous intrinsic properties.   

Our analysis is not well suited
to the two cluster lenses, RXJ0911+0551 and Q0957+561, because our use
of an external shear to represent perturbations is breaking down.  
If, however, we assume that the true surface density of the primary
lens is the same as for the other lenses, we can estimate the external
convergence $\kappa_{ext}$ required to make the time delay estimates for
the cluster lenses agree with the other 5 systems.  We find that
$\kappa_{ext,0911} \simeq 1.11-1.43h \pm 0.10$ and 
$\kappa_{ext,0957} \simeq 0.99-0.45h \pm 0.01$.
While these results should not be taken 
at face value because of the problems with our approximation for
lenses which are too close to the central regions of clusters, the 
pattern of the solutions seems reasonable. The RXJ0911+0551 lens 
lies $200h^{-1}$~kpc from the cluster center, while the Q0957+561
lens lies only $30h^{-1}$~kpc from the cluster center so we would
expect larger surface densities in Q0957+561 than in RXJ0911+0551.   
Presumably similar analyses with more realistic models for the cluster
would improve matters.

Our semi-analytic models for the four simple lenses with good photometry,
PG1115+080, SBS1520+530, B1600+434 and HE2149--2745, almost exactly 
reproduce the $H_0$ estimates based on full numerical models by  
Kochanek~(\cite{Kochanek02a}). We find $H_0\simeq 51\pm 5$~km/s~Mpc
if $\kappa=1/2$ and $\eta=2$ as expected in dark matter dominated lenses
and $H_0 \simeq 73\pm 8$~km/s~Mpc if they have constant mass-to-light
ratios.  The dependence of the time delays on the radial and angular structure of the
lens shows that standard parametric models, which can adjust both the radial
mass distribution and the quadrupole structure of the lens, encompass the physical 
properties needed to study the dependence of Hubble constant estimates on
the mass distribution of the lens.  Most degrees of freedom in non-parametric
models (e.g. Williams \& Saha~\cite{Williams00}) are not important for time
delay estimates in simple lenses.  Since the parametric models have the 
advantage of corresponding to physical models of galaxies, while most 
realizations of the non-parametric models do not (see Munoz et al.~\cite{Munoz01}),
the use of non-parametric approaches is probably better suited to very complicated
systems like B1608+656 where the lens consists of two interacting galaxies 
(Koopmans \& Fassnacht~\cite{Koopmans99}) and it is unclear how to properly
parameterize the system.  It is also easy to make theoretical estimates for
the time delays expected for standard cold dark matter (CDM) halo models,
since we need only estimate the surface density $\kbar$ of the halo.  We
examine this problem in Kochanek~(\cite{Kochanek02b}).

\noindent Acknowledgments.  
CSK thanks D. Rusin, P. Schechter, J. Winn and S. Wyithe for discussions and comments. 
CSK is supported by the Smithsonian Institution and NASA ATP grant NAG5-9265.

\end{document}